\documentclass[letterpaper]{emulateapj}

\def \apj 		{Astrophys. J.}

\def \aj 		{Astron. J.}
\def \aap 		{Astron. \& Astrophys.}

\def \mnras 		{Month. Not. Royal Astron. Soc.}

\def \kms		{ km s$^{-1}$ }			
\providecommand{\msol}	{\ensuremath{M_{\odot}}}	

\def \eqtext#1		{\hspace{1in} \hbox{#1}}	
\providecommand{\micron}{\ensuremath{\mu\rm{m}}}

\def \etal      	{{\it et al.}\ }		

\def \idlplot#1		{\centerline{\scalebox{0.9}{\includegraphics{#1}}}} 
\def \idlplotps#1	{\centerline{\scalebox{0.9}{\includegraphics[70,350][574,710]{#1}}}} 
\def \ie		{{\it i.e.\/}}
\def \eg		{{\it e.g.\/}}
\def \subsimt#1{{\lower 2pt\hbox{$\scriptstyle #1$}\atop
     \raise 1pt\hbox{$\scriptstyle \sim$}}}

 1
 2

 1
 2

\usepackage{natbib}
\received{November 29, 2004}
\revised{February 21, 2005}
\accepted{March 16, 2005}
\shorttitle{Photo-evaporation and planetesimals}
\shortauthors{Throop \etal}
\begin{document} 
\bibliographystyle{apj}
\title{Can photo-evaporation trigger planetesimal formation?}

\author{Henry B. Throop}
\affil{Department of Space Studies}
\affil{Southwest Research Institute}
\affil{1050 Walnut St, Ste 400, Boulder, CO  80302}
\email{throop@boulder.swri.edu}

\author{John Bally}
\affil{Center for Astrophysics and Space Astronomy}
\affil{University of Colorado}
\affil{UCB 389, Boulder, CO 80309-0389}

\begin{abstract}
We propose that UV radiation can stimulate the formation of planetesimals in
externally-illuminated protoplanetary disks.  We present a numerical model of
disk evolution including vertical sedimentation and photo-evaporation by an
external O or B star.  As solid material grows and settles toward the disk
midplane, the outer layers of the disk become dust depleted.  When such a disk
is exposed to UV radiation, heating drives photo-evaporative mass-loss from its
surface, generating a dust-depleted outflow.  The dust:gas surface density
ratio in the disk interior grows until dust in the disk midplane becomes
gravitationally unstable.  Thus, UV radiation fields may induce the rapid
formation of planetesimals in disks where sedimentation has occurred.
\end{abstract}

\keywords{solar system: formation -- (stars:) planetary systems: protoplanetary disks -- (stars:) planetary systems: formation}

\section{Introduction} 
\label{sect:intro}

The majority of stars are born from giant molecular clouds in dense but
transient OB associations such as Orion.  Nearby O and B stars can
photo-evaporate the outer parts of circumstellar disks surrounding low-mass
stars on $10^5-10^6$ yr timescales.  Several dozen disks with diameters from
100~AU to 1200~AU have been directly imaged around young stars embedded in the
Orion nebula with the Hubble Space Telescope (HST) \citep{ow96,bsd98,bom00}.
Excess near-infrared emission in the 2 to 10 $\mu$m wavelength region
\citep{sbs04} provides indirect evidence that more than 80\% of the several
thousand low-mass young stars in the Orion nebula cluster are surrounded by
disks of gas and dust.  Disks which are not directly seen in images are either
too small, are lost in the glare of their central stars, or are hidden inside
externally ionized cocoons (`proplyds') produced by their own
photo-evaporation.  About half of these disks are being photo-evaporated by
ultraviolet (UV) radiation from nearby O and B stars in the Trapezium cluster.
Mass-loss rates have been measured to be in the range $10^{-7} - 10^{-6} \msol\
\rm{yr^{-1}}$ \citep{ho99}, implying the loss of a disk within $10^4$ to $10^6$
years.  It has been suggested that the formation of planets in regions such as
Orion may be difficult -- planets must form either rapidly, or rarely
\citep{tbe01,hdh04}.

In this letter, we use a numerical model to study the loss of material from
a photo-evaporating disk and to examine the effects of sedimentation towards
the disk midplane on the loss of the gas and dust components.  Previous models
of disk evaporation \citep{jhb98,mjh03,ahl04} have examined only the gas
component, assuming that gas and dust are well-mixed throughout and that the
loss of gas implies the loss of dust.  In our model, the evolution of the gas
and dust components are treated individually.  We explore the consequences of
grain growth and sedimentation on planetesimal formation.  Our model assumes
that planetesimal formation begins by non-gravitational sticking in collisions
\citep[\eg][]{cth93}, after which material sediments to the midplane where it
can undergo gravitational instability after sufficient gas is removed
\citep{gw73,ys02}.  The model presented here builds on our earlier work that did
not include vertical sedimentation or gravitational instability
\citep{thr00,tbe01}.  


%
%
%
%

\section{Disk Model} 
\label{sect:model}

We have developed a numerical model which tracks the state of the disk's gas
and dust components as they undergo vertical sedimentation and external
photo-evaporation.  We consider the effects of UV-induced photo-evaporation due
to nearby massive stars, and we ignore the effects of self-irradiation by UV
light from the low-mass stars embedded within the disks.  Thus, our model is
most applicable to the outer portions of disks, or to disks which are
effectively shielded from self-irradiation.  

Most disks around low-mass stars in OB associations formed well before
photo-evaporation began.  Low-mass stars in Orion have ages $10^5 - 10^6\
\rm{yr}$ \citep{hil97}, while most appear to have been photo-evaporating for
only a few $10^4\ \rm{yr}$ \citep{bom00}.  Therefore, disks can evolve
quiescently as grains grow and settle for up to about $10^6\ \rm{yr}$, after
which photo-evaporation begins.

\emph{Initial Disk Structure.} The initial surface density follows the model of
\citet{hay81}.  Surface densities for the gas and dust as a function of
distance $r$ are set by:

\begin{equation}
\Sigma_d = 30.1\  (r/{\rm{AU}})^{-k}\  [\rm{g\ cm^{-2}}]
\end{equation}

\begin{equation}
\Sigma_g = 1700\  (r/{\rm{AU}})^{-k}\ [\rm{g\ cm^{-2}}],
\end{equation}
implying a dust:gas mass ratio of 1:60.  We assume a mass distribution exponent
$k = 3/2$ and a disk size $r_d = 100\ \rm{AU}$, yielding a disk mass
$0.02~\msol$, with half of this between 1~AU and 30~AU.   In the vertical
direction $z$, we use the \citet{hay81} pressure-balance
scale height $H = 7.1 \times 10^{11} ({r/ \rm{AU}})^{-5/4}\ \rm{cm}$, and
calculate the gas density
\begin{equation}
\rho_g(r,z) = {\Sigma(r) \over {2 H}} e^{-({z/H})^2} \ [\rm{g\ cm^{-3}}],
\end{equation}
where the factor of two accounts for the disk's top and bottom sides.  The
central star has 1~\msol.

\emph{Grain growth and sedimentation.} Dust grains in young disks are
transported by small-scale turbulent eddies.  Within these eddies, grains
collide, stick, and grow \citep{cdh96}; as they grow, they sediment toward the
disk midplane.  Numerous studies have established that most of the disk's dust
mass will have grown via collisions to sizes of cm or beyond and sedimented to
the midplane within a few $10^4-10^5\ \rm{yr}$. \citet{wei97} found that by
$7\times10^4\ \rm{yr}$, typical bodies at 30~AU had grown to meter sizes and
settled to the midplane.  \citet{mmv88} found that typical grains at 40~AU grew
to nearly cm-sizes after 2400~yr, and \citet{thr00} found that grains grew to
cm sizes at 100~AU within $10^5\ \rm{yr}$.  HST observations directly support
grain growth, even at the 600~AU outer edge of Orion's largest known disk
\citep{tbe01}.  Assuming a settling time $t_s \simeq 10^6\ (r/\micron)^{-1}\ \rm{yr}$
\citep[\eg,][]{ys02}, a typical disk will easily be settled to beyond 100~AU in
the $10^6\ \rm{yr}$ for which it can evolve before the onset of
photo-evaporation.  


\emph{Vertical structure.} Local turbulence will prevent grains from settling to an
infinitely thin plane.  The equilibrium vertical distribution of dust in a
proto-planetary disk has been studied by \citet{sek98}.  In their model, the density
is determined by an equilibrium between grain settling and grain lofting via
the Kelvin-Helmholtz `sandstorm' instability.  Because the gas disk has only a
finite mass-loading capacity, the dust density increases rapidly toward the
midplane.  In some cases, the solution for midplane density becomes singular.
\citet{ys02} interpreted this singularity as being indicative of the dust
disk's susceptibility to a gravitational instability (GI).  They did not model
planetesimal formation from this instability, only the onset of the GI.  They
found that in the \citet{hay81} disks, an increase in the dust:gas surface
density of $\sim 2-10$ (varying with $r$) was sufficient to reach this
instability, and it could be caused either by increasing $\Sigma_d$ or
decreasing $\Sigma_g$.

The Sekiya vertical profiles assume that the disks have no global turbulence.  This
condition has not been established in the Orion disks; indeed, many models describing
turbulence from thermal convection and magneto-hydrodynamic instabilities have been
published \citep[\eg,][]{lp80,bhs96}.  Because of their proximity to luminous sources
of radiation, however, Orion's disks are expected to be relatively warm with
temperatures of around 50~K.  Thus, turbulence produced by thermal convection from
the disk mid-plane is expected to be suppressed compared to that in similar disks in
dark clouds such as Taurus where the disks may be as cold as 10~K.  However, the
precise nature of any turbulence remains an open issue.


Observations provide indirect evidence that the Orion disks have settled.
\citet{jhb98} estimate from model fits that the disks' outer shells are
characterized by a UV penetration depth $N_D = 3\times 10^{21}\ \rm{cm^{-2}}$
-- that is, the outer shell is depleted in dust by at least a factor of $3$
from the classical $N_D$.  This would be expected in a disk in which solids had
partially sedimented to the midplane.

\emph{Photo-evaporation.}  UV radiation heats the disk surface to a depth $N_D$
where the radiation is predominantly absorbed by dust.  Wherever the sound-speed
in the heated layer exceeds the local gravitational escape speed, mass-loss
occurs.  Mass-loss occurs in two distinct regimes \citep{jhb98}.  FUV radiation
($912 \rm{\AA}\ < \lambda < 2,000\rm{\AA}$) produces a warm $10^3\ \rm{K}$
neutral hydrogen outflow for radii $r_I > GM / 2 c_I^2$ where $c_I \sim 3$ \kms\
is the sound-speed, and the factor of roughly two accounts for the effects of
pressure gradients within the outflow.  For a 1 \msol\ star, $r_I \sim 50$ AU.
For most disks, the resulting mass loss prevents ionizing EUV radiation
($\lambda < 912$ \AA ) from reaching the disk surface.  But, as a disk loses
mass and shrinks to a size less than $r_I$, FUV-induced mass-loss declines and
stops entirely.  The ionizing EUV radiation can then reach the disk surface,
ionize its skin, and raise its sound-speed to $c_{II} \sim 10\ \rm{km\ s^{-1}}$.   
Mass loss resumes in the form of a fully ionized wind until a radius of
about $r_{II} \sim GM / 2 c^2_{II} \sim $5 AU.

We incorporate the mass loss rates of \citet{jhb98} from their eqs. (5) and
(23).  We assume an outflow wind speed $v_0 = 3\ \rm{km\ s^{-1}}$, and a UV
penetration depth $N_D = 3\times10^{21}\ \rm{cm}^{-2}$ (\ie, their $\epsilon =
3$).  We assume that small, micron-sized dust grains produced as collisional
byproducts will remain well-mixed with the gas and therefore continue to provide
efficient heating of the gas as they absorb UV radiation, even as larger grains
settle to the midplane.  The external star is assumed to have an output of
$10^{49}\ \rm{photons\ s^{-1}}$ and be at a distance $d = 0.1\ \rm{pc}$, 
roughly corresponding to Orion's 182-413 (HST10) disk.

Following \citet{jhb98}, we assume that the mass-loss rate per area is constant
across the disk.  We assume that mass is lost from the top down (decreasing
$z$), and that the mass loss occurs symmetrically on both sides of the disk.  If
the disk material re-settles vertically on a timescale faster than the
photo-evaporation time, then the disk will maintain a general symmetry in the
$z$ direction, regardless of asymmetry in the illumination source.  This
condition is met (for dust and gas respectively), if

\begin{equation}
{\Sigma(r) \over {\dot \Sigma}} > t_{\rm s}(a(r)) 
\label{eq:symmetrydust}
\end{equation}
and
\begin{equation}
{\Sigma(r) \over {\dot \Sigma}} > {H(r) \over v_0} \ .
\end{equation}
Using typical grain sizes $a(r)$ due to coagulative growth, and settling times
$t_{\rm{set}}$ \citep[\eg][]{mmv88,wei97,thr00}, we find that these conditions
are satisfied throughout our simulation.

\emph{Dust entrainment.}  When gas is lost to the UV-induced outflow, small
grains are entrained and dragged along with the flow. We assume, by analogy with
the gas's finite mass-loading \citep{ys02}, that gas can entrain
its own mass density of dust grains, but no more.  The entrained dust
volume-density in the outflow is therefore $\rho_{de} = \rm{Min}(\rho_d(r,z),
\rho_g(r,z))$.  In most cases, these densities are equal near roughly $H/100$,
causing a preferential removal of gas above this location.
 

\emph{Model structure.} Our numerical model uses a 100$\times$100 2D grid to
track the densities $\rho_g(r,z)$ and $\rho_d(r,z)$.  We use
logarithmically-spaced bins in the vertical and radial directions, and assume
azimuthal symmetry.  The model uses a self-adjusting variable timestep; typical
steps are $10^3\ \rm{yr}$ at the outer edge and $10\ \rm{yr}$ at the inner.
The timesteps chosen are more than sufficient to assure numerical convergence.
At each timestep, photo-evaporation and grain entrainment act on the disk as
described above.  Our simulation begins when the disk first becomes exposed to
UV radiation from nearby massive stars; it ends when the disk has been eroded
in to $r_d \leq r_{II}$, stopping all photo-evaporation.

\section{Results} 

\label{sect:results}

\begin{figure}[ht]
\centerline{
  \scalebox{0.50}{ \includegraphics{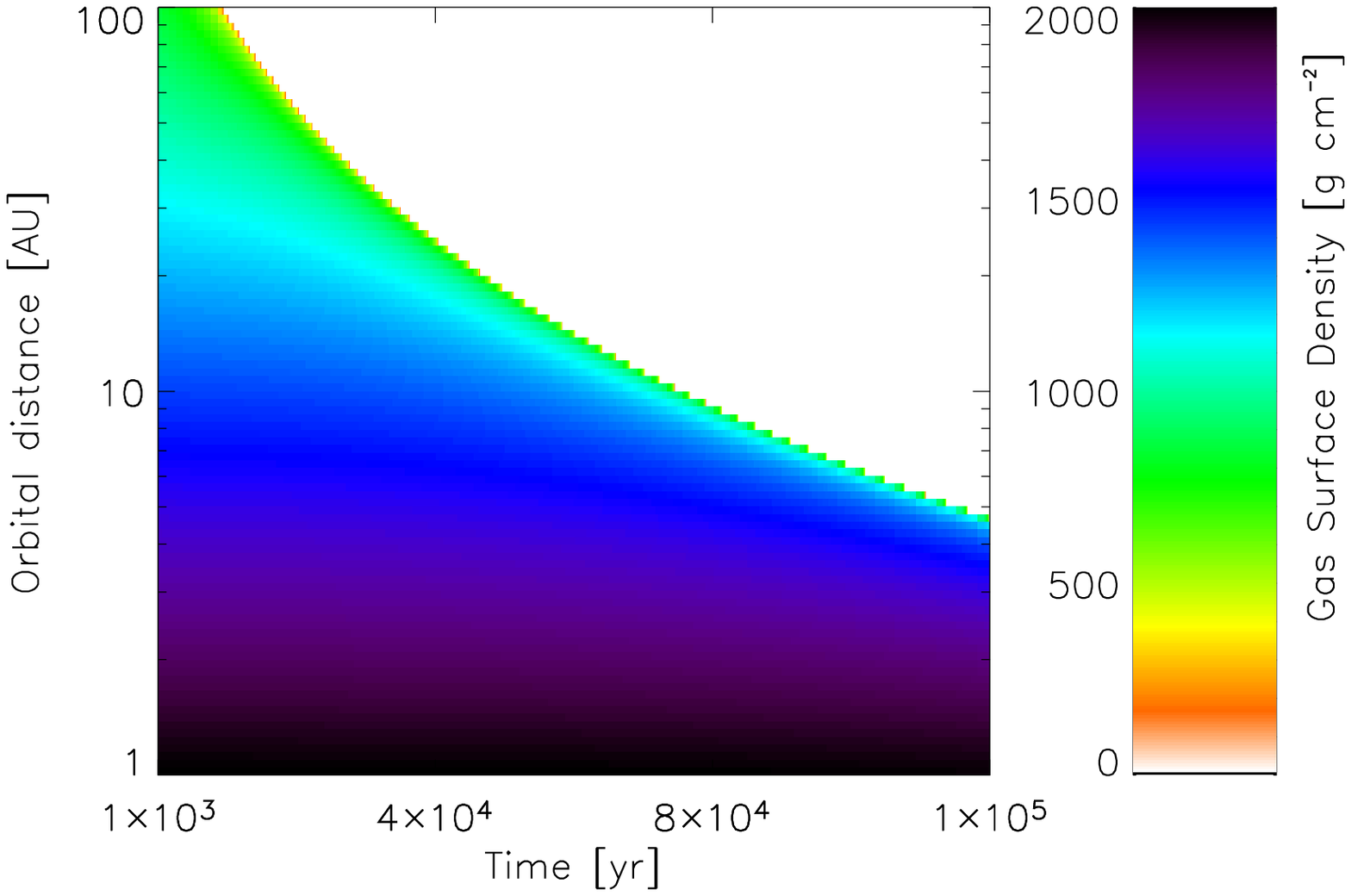}}
  \scalebox{0.50}{ \includegraphics{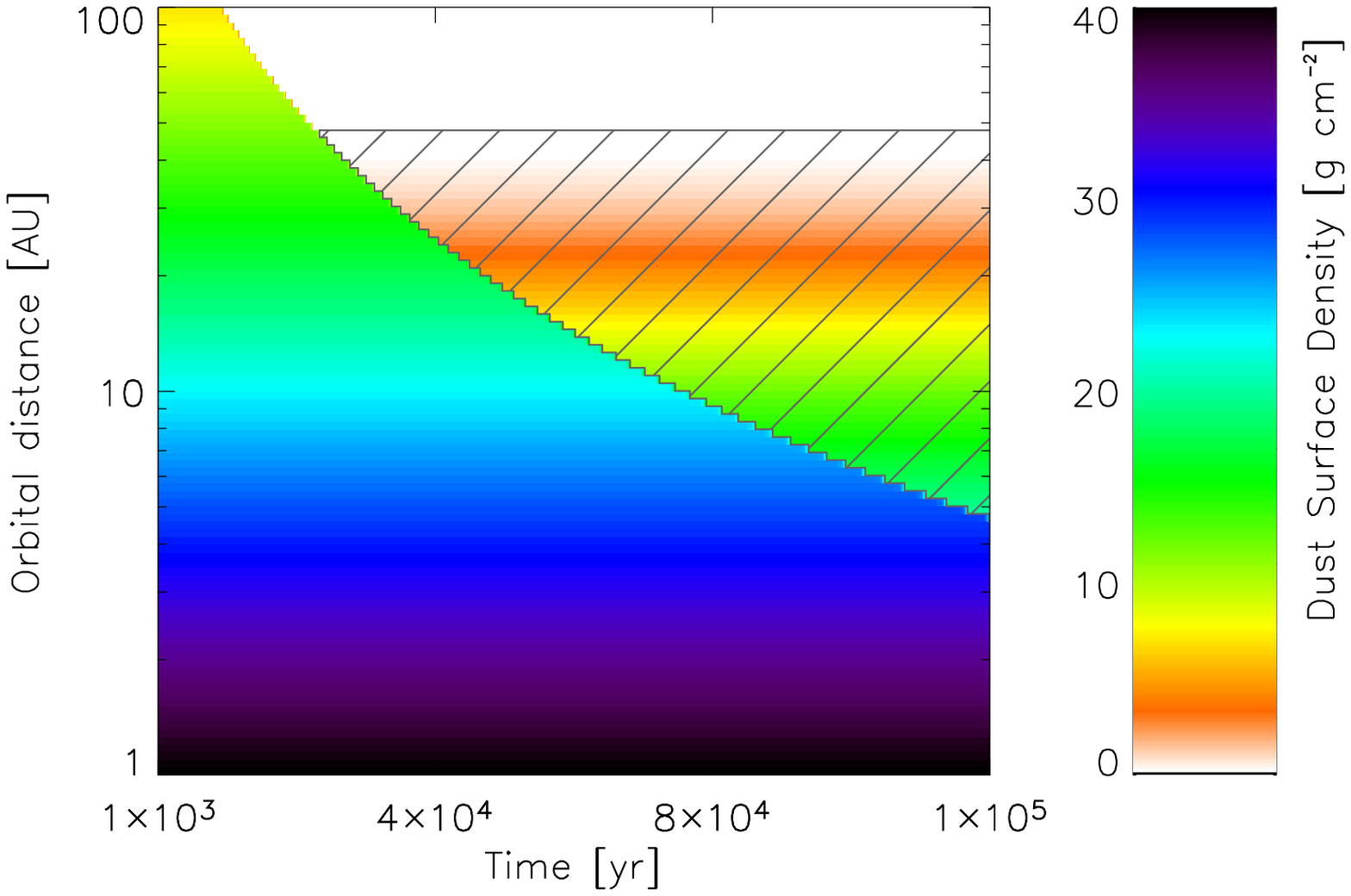}}
}
\caption{Evolution of our model circumstellar disk of mass 0.02~\msol.  Before
the model begins, dust grains have grown and settled toward the midplane via
Sekiya's vertical distributions.  Photo-evaporation fully removes gas inward to
5~AU (left), while leaving substantial amounts of dust in this same region
(right). The dust:gas surface density ratio through much of this region
increases to a level where the dust is unstable against gravitational collapse
(hatched region, right).} 
\label{fig:sekiya} 
\end{figure}

Figure \ref{fig:sekiya} shows the output from our model run.  The disk is
gradually eroded from the outside edge inward, first by FUV radiation and then
by EUV.  After $1.25\times10^{5}\ \rm{yr}$, the gas disk has been removed
entirely outward of 5~AU, and continued UV flux will remove no additional gas
inward of this distance.  Dust throughout the disk is partially but
not fully removed: between 5 and 50~AU, the disk is left with roughly 4~Earth
masses of dust, or 3\% of its original dust mass.  This dust has condensed to
the midplane and is not entrained in the photo-evaporative outflow.  It meets
the instability criteria of \citet{ys02}, indicating that km-scale
planetesimals could spontaneously form.  Beyond 50~AU the dust density is not
high enough to meet the instability criteria.  Between 1 and 5~AU, an
additional 20 Earth masses of solid material remains; this material is
unaffected by external photo-evaporation and thus although it has settled, does
not become unstable because gas motions inhibit collapse.

We have performed a second run on a disk with all of the same characteristics,
except that the gas and dust densities $\Sigma_g$ and $\Sigma_d$ were increased
to three times their nominal values, giving a disk mass of $0.06\msol$.  In this
run, the general behavior was similar to that shown in our nominal run.
However, between 5 and 50~AU nearly 140~Earth masses of material remained, or
40\% of the original dust mass.  The entire remaining disk met the instability
criteria.  The highly nonlinear dependence between the disk mass and the
fraction of dust retained is due to the fact that nearly all the additional mass
goes directly to the midplane.  Initial results indicate that this trend
continues with even higher dust masses.

\section{Discussion} 
\label{sect:discussion}

If a disk has more than a few $10^5$ years to evolve before it is exposed to UV
radiation, grains throughout the disk can grow sufficiently large to settle
toward the disk mid-plane.  A gradient in the vertical distribution of the
dust:gas ratio is established, and the disk surface layers become dust-depleted
while the disk mid-plane becomes dust-enriched.  Because photo-evaporation
preferentially removes the gas component, the settled dust subdisk remains and
the gas no longer inhibits a gravitational instability.  The subsequent
formation of planetesimals by gravitational instability addresses a size scale
(cm to km) that has been difficult for purely accretional growth models, because
bodies in this size range have neither the self-gravity nor the internal
strength to maintain integrity after collisions of more than several $\rm m\
s^{-1}$ \citep[\eg,][]{wc93}.

%
%
 
Planetesimal formation by means of gravitational instability may be enhanced by
any process that sufficiently concentrates dust.  For example, \citet{ys02}
presented another method for enhancing the dust:gas ratio -- the inward radial
drift of cm-to-meter-scale  bodies in the nebula.  They demonstrated that local
instabilities could be created on timescales of $10^5$-$10^6\ \rm{yr}$ in the
1-100~AU region.  The methods presented in both our work and theirs operate on
roughly comparable distances and timescales.

Our model is simplistic and ignores some important processes.  \citet{hjl94}
found that the central star's own UV radiation can drive photo-evaporation.
Because we have ignored such self-irradiation in our models, the results
presented here apply to the outer regions of disks (beyond several~AU), or to
disks shielded from illumination from their central stars.  Self-illumination
acts in a similar way to external illumination and will also preferentially
remove gas, so we expect that the effect in the inner region will be similar to
that demonstrated here.  Additionally, the inner disk will be affected by
viscous transport of material toward the central star.  Our model does not
include these processes; \citet{mjh03} found that self-irradiation and viscous
transport together formed gaps near the inner gravitational radius.  Such gaps
may halt the inward radial drift of solid material, further spurring the onset
of instability at that location.

OB associations have been accused of being hazardous to planetary formation
\citep[\eg][]{tbe01,hdh04}.  On the contrary, our results indicate that UV
radiation fields in an OB association may actually stimulate a critical step in
planet formation -- the growth of kilometer-size planetesimals.  If our Solar
System formed via the method described here, the initial disk mass must have
been greater than 0.02\msol\, because that disk leaves insufficient material
outward of 5~AU from which to form the giant planet cores, assuming the
standard core-and-envelope scenario of \citet{phb94}.  These planets have cores
totaling roughly 40~Earth masses; our simulation leaves only a total of 4 Earth
masses.  If the disk were only several times more massive, however, the
nonlinear dependence between initial and final dust masses would imply that
sufficient mass for the cores would remain.  The formation of planetesimals in
our model depends on the loss of most of the disk's gas envelope, after which
there would be insufficient material from which to form the giant planets'
atmospheres; therefore, if giant planets form near OB associations, they may
have formed by more rapid methods \citep[\eg,][]{bos03}.  Partial
retention of the gas envelope enabling giant planet formation could occur if
photo-evaporation ceases due to the short lifetimes of the illuminating stars,
or if the gravitational radius $r_{II}$ were moved outward from 5~AU.

Our model may underestimate the amount of solid material left in the disk, for
two reasons.  First, as grains continue to settle out, the dust:gas ratio in the
outer shells where photo-evaporation occurs will decrease, causing a
corresponding decrease in the amount of dust entrained.  Because $N_D$ is
considered a constant in our model, we do not account for this decrease in the
fractional dust-loss rate.  Second, our dust entrainment criterion has no grain
size dependence.  In reality, as grains grow they will become more difficult to
entrain, and our method will overestimate their loss rate.  In both cases, more
detailed modeling of the gas-grain interaction is necessary; we estimate that
after these factors are included, the disks may retain up to several times more
material than indicated here.  A future paper will present a more complete model
of grain growth and entrainment.  Additional work remains to be done to examine
the stability and evolution of volatiles in these systems.

\section{Acknowledgments} 

This work was supported by SwRI IR\&D program 15.R9447 and NASA Astrobiology
grant NCC2-1052.  We thank W. Ward, D.  Nesvorn\'y, S. A. Stern, and reviewer
S. Weidenschilling for their useful comments.

\end{document}